\documentclass[a4paper]{jpconf}
\usepackage{graphicx}
\usepackage{epsfig} 
\begin{document}
\title{ArgoNeuT and the Neutrino-Argon Charged Current Quasi-Elastic Cross Section}

\author{Joshua Spitz, for the ArgoNeuT Collaboration}
\address{Department of Physics, Yale University, New Haven, CT 06520, USA}
\ead{joshua.spitz@yale.edu}

\begin{abstract}
ArgoNeuT, a Liquid Argon Time Projection Chamber in the NuMI beamline at Fermilab, has recently collected thousands of neutrino and anti-neutrino events between 0.1 and 10~GeV. The experiment will, among other things, measure the cross section of the neutrino and anti-neutrino Charged Current Quasi-Elastic interaction and analyze the vertex activity associated with such events. These topics are discussed along with ArgoNeuT's automated reconstruction software, currently capable of fully reconstructing the muon and finding the event vertex in neutrino interactions. 
\end{abstract}

\section{The LArTPC technique}
\begin{figure}[h!]
\centering
\begin{tabular}{c}
\epsfig{file=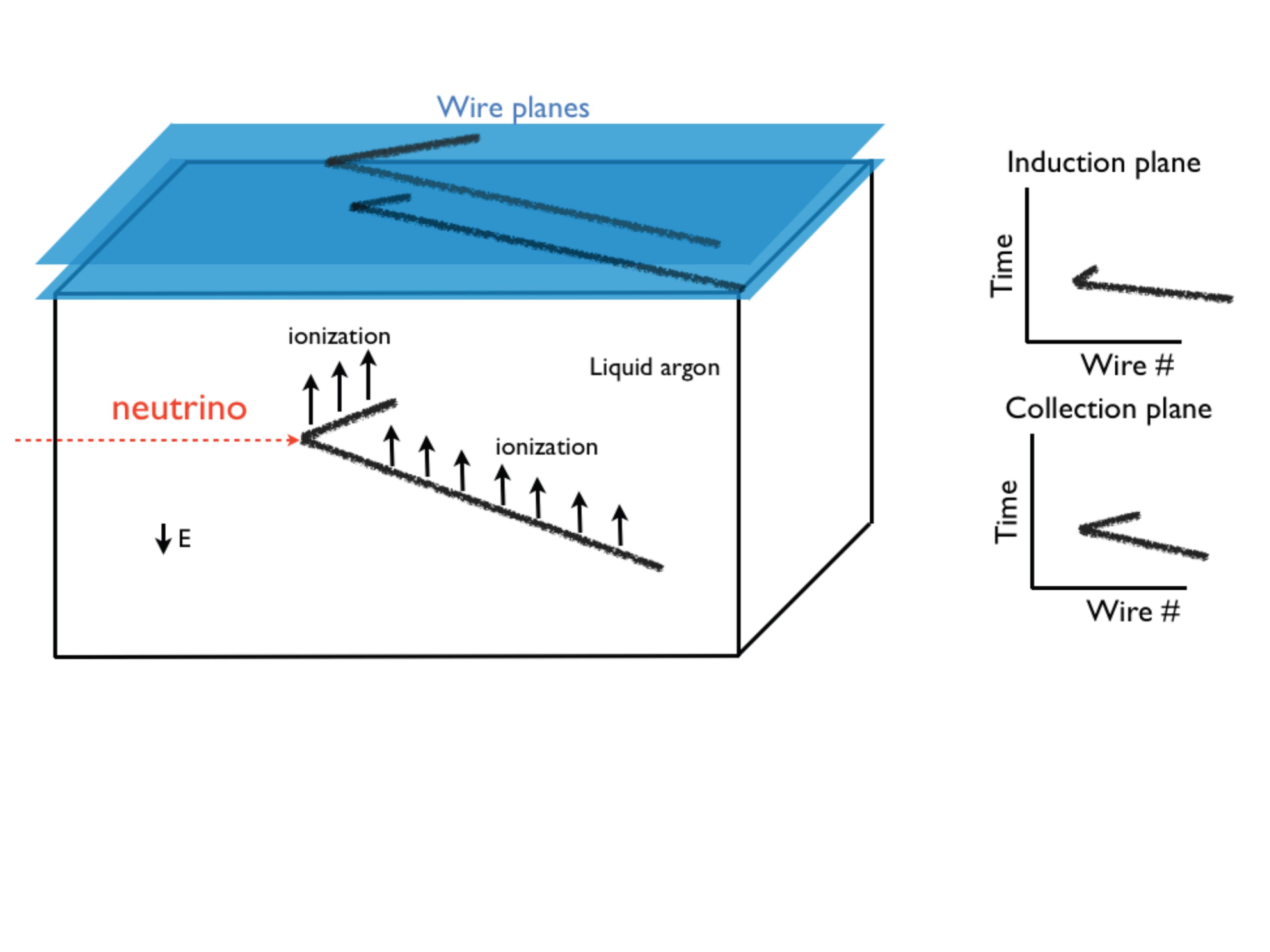,width=4.9in,angle=0}
\end{tabular}
\vspace{-2.5cm}
\caption{The LArTPC concept for neutrino detection.}
\label{lartpc}
\end{figure}

Liquid Argon Time Projection Chamber (LArTPC) technology~\cite{rubbia} provides mm-scale position resolution, calorimetry, three dimensional imaging, and efficient particle identification for neutrino detection and reconstruction. In a LArTPC, the interaction products exiting the neutrino vertex ionize argon atoms as they traverse the detector. As argon is a noble element, this ionization is free to drift across the medium. An electric field is imposed in the detector and the ionization tracks are drifted towards readout wire planes that are oriented with respect to one another at an angle. The ionization induces a current on the induction plane and is collected by the collection plane. Combining the information from both planes and detecting the events in time allows a three dimensional image of the event with calorimetric information to be obtained. Argon, with a relatively large density, short radiation length, and high scintillation yield, is ideal for neutrino detection and event containment. A schematic depicting the LArTPC concept for neutrino detection can be seen in Figure~\ref{lartpc}. 

\section{ArgoNeuT}
\begin{figure}[h!]
\centering
\begin{tabular}{c c}
\hspace{-1.8cm}
\epsfig{file=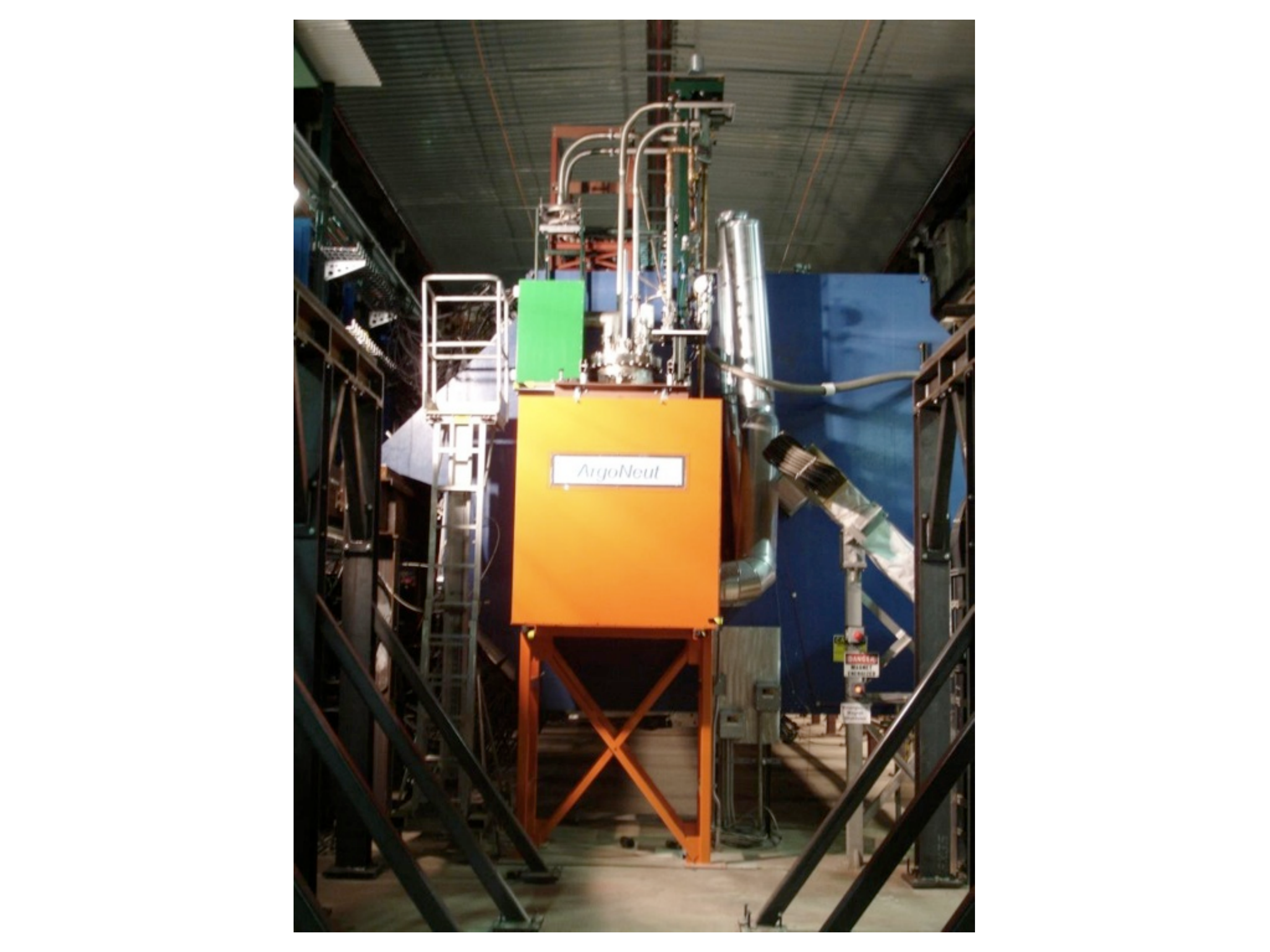,width=4.4in,angle=0}
&
\hspace{-4.cm}
\epsfig{file=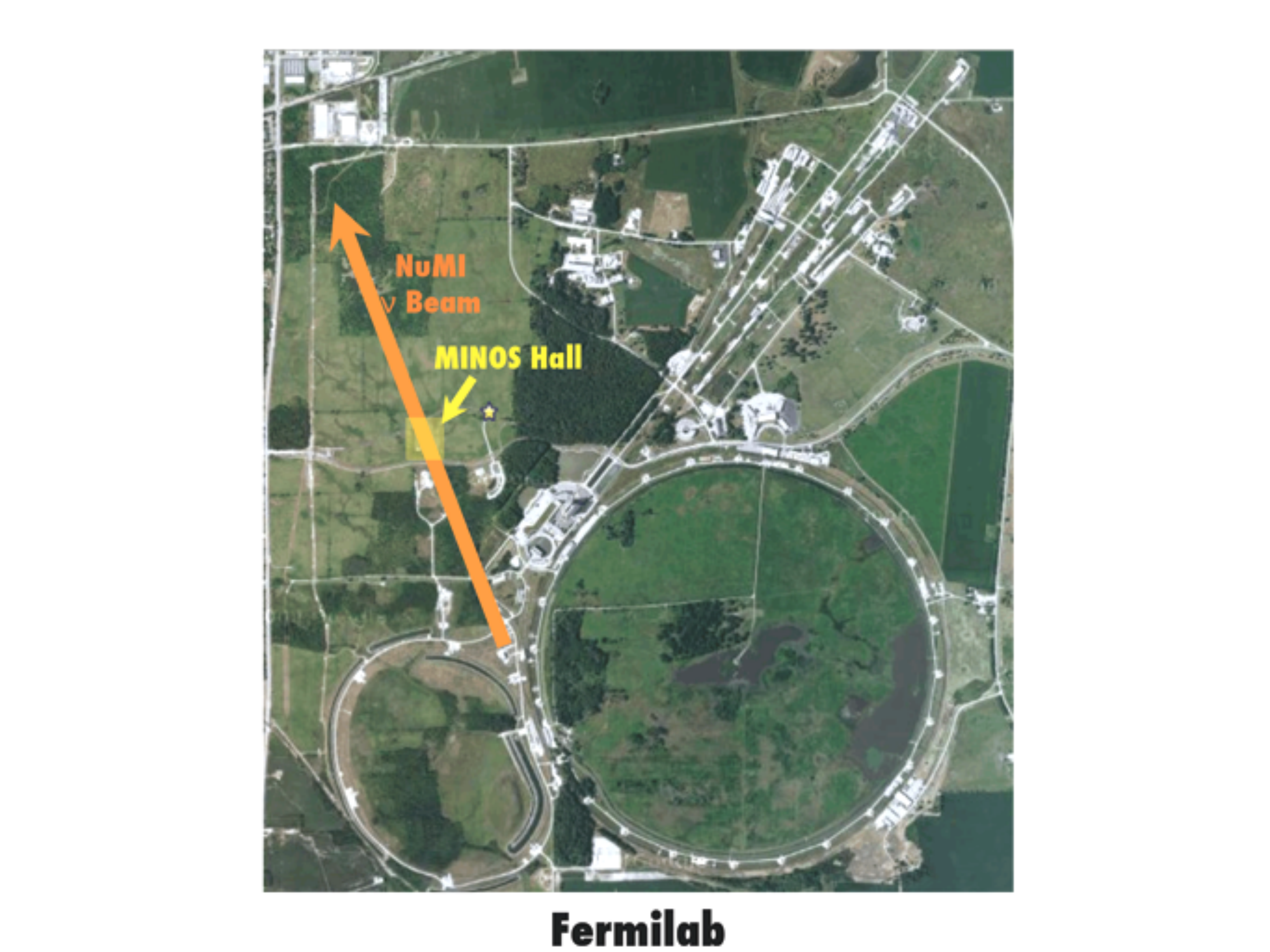,width=4.4in,angle=0}
\end{tabular}
\vspace{0.0in}
\caption{(Left) ArgoNeuT sitting just in front of the MINOS near detector in the NuMI beamline. (Right) The location of ArgoNeuT in the MINOS hall, relative to the Fermilab accelerator complex.}
\label{detectorpic}
\end{figure}

ArgoNeuT (NSF/DOE) sat just upstream of the on-axis MINOS~\cite{MINOS} near detector in the NuMI beamline, about 1~km from the target station and 100~m underground at Fermilab (see Figure~\ref{detectorpic}). The experiment features a 170~L active volume TPC with a fully contained recirculation and purification system. The main detector characteristics are summarized in Table~\ref{vitals}. ArgoNeuT collected about 1.35E20 protons on target (POT) with the ``low energy" NuMI configuration (0.1E20 POT in neutrino mode and 1.25E20 POT in anti-neutrino mode) from September 2009 to February 2010. Run operations were largely stable and shift-free over the $>$5 month span. Along with a Charged Current Quasi-Elastic (CCQE) cross section measurement (one of many potential cross section measurements) ArgoNeuT will be demonstrating the dE/dx and topological-based particle identification capabilities of LArTPCs with a special emphasis on electron/photon differentiation, a vital step toward the realization of a future background-free electron-neutrino appearance search. ArgoNeuT is also developing reconstruction techniques specific to LArTPCs in collaboration with MicroBooNE~\cite{microboone} and LBNE~\cite{lbne} through the ``LArSoft" collaboration and performing research and design for future LArTPCs. Indeed, ArgoNeuT is one of the key steps in the phased US LArTPC program moving toward construction of a $>$10~kiloton detector to be used for long-baseline and atmospheric neutrino oscillation physics, a proton decay search, supernova burst and diffuse neutrino detection, and more~\cite{modular,landd,glacier,flare,icanoe2,Conrad:2010mh,Cocco:2004ac,Rubbia:2004yq,Arneodo:2001tx}.

\begin{table}[h]
\centering
\begin{tabular}{|c|c|}
	\hline
Cryostat volume  &500 liters  \\ \hline
TPC volume  &175 liters  \\ \hline
\# Electronic channels  &480 (240/plane)  \\ \hline
Wire spacing  &4 mm  \\ \hline
Electronics style (temperature)  &JFET (293 K)  \\ \hline
Max. drift length (time)  &50 cm (330 $\mu$s)  \\ \hline
Light collection?  &No  \\
\hline
\end{tabular}
\caption{The ArgoNeuT detector's main specifications.}
\label{vitals}
\end{table}
\vspace{-.5cm}

\section{The CCQE cross section and vertex activity}
The Charged Current Quasi-Elastic (CCQE) interaction is the most consequential neutrino interaction channel for accelerator-based neutrino oscillation experiments. The $\nu_{\mu}$ ($\overline{\nu}_{\mu}$) CCQE interaction $\nu_{\mu}n\longrightarrow\mu^{-}p$ ($\overline{\nu}_{\mu}p\longrightarrow\mu^{+}n$) is used to measure the muon-neutrino flux, constraining the expected electron-neutrino flux at the near and far detectors. Long baseline electron neutrino appearance searches (MINOS~\cite{MINOS}, T2K~\cite{t2k}, NOvA~\cite{nova}, LBNE~\cite{lbne}, etc.) all employ a near and far detector in order to minimize neutrino flux uncertainties. However, due to the non-zero near-far detector systematics (i.e. differences between the near and far detector) inherent in all long baseline experiments, cross section assumptions need to be made\footnote{Such systematics come from the differences in the neutrino energy spectra seen at the near and far detector. Also, systematics can come from inherent differences in detector parameters (detector medium, size, acceptance, etc.) between near and far.}. Knowledge of the neutrino cross section is required even in the case of a near-far configuration.  The simple two particle event topology and dominant cross section in the most relevant energy region makes the CCQE interaction the golden channel for long baseline neutrino oscillation experiments. Despite being the golden channel, however, the CCQE cross section is known with only 20-30\% precision over most energies. Furthermore, only one (preliminary) measurement of the CCQE cross section on argon has ever been taken (at $<E_{\nu}>$=28~GeV)~\cite{icaruswanf,wanfprd}.  

The neutrino energy can be reconstructed with the lepton energy and angle with respect to beam axis alone in the case of a CCQE interaction. LArTPCs have the added advantage of the ability to reconstruct the proton(s) that come out of the interaction and, with knowledge of the lepton kinematics, fully reconstruct the neutrino energy. The additional kinematic constraints from the reconstructed proton(s) aid in neutrino energy resolution and background suppression. The muon and proton(s) are identified unambiguously in a $\nu_{\mu}$ CCQE event with a combination of dE/dx and range measurements in ArgoNeuT. The downstream magnetized MINOS near detector is used to sign-select and, with ArgoNeuT, fully reconstruct the muon while the proton(s) (contained $>$50\% of the time) is reconstructed with ArgoNeuT alone. 

CCQE cross section measurements are subject to large systematics coming from final state interaction (FSI) modeling, flux uncertainties, background neutrino charged current pion interaction cross section uncertainties, and more. Charged current pion events ($\nu_{\mu}p\longrightarrow\mu^{-}p\pi^{+}$, $\nu_{\mu}n\longrightarrow\mu^{-}p\pi^{0}$, or $\nu_{\mu}n\longrightarrow\mu^{-}n\pi^{+}$) can appear ``CCQE-like" if the pion is absorbed (e.g. $\pi N \rightarrow NN$) during an interaction within the nucleus. More CCQE backgrounds can come from interactions involving final state neutral particles (neutrons, gammas) that escape detection and/or neutral current events with a final state pion incorrectly identified as a muon that renders the event ``CCQE-like". In true CCQE events the nucleon is also subject to various FSI processes including pion production, charge exchange, and elastic and inelastic scattering within the parent nucleus. Flux-integrated differential cross sections with muon kinematics and cross sections based on post-FSI observable particles only (rather than pre-FSI, which introduces a model dependence in an attempt to track back to the neutrino interaction itself) are now starting to be reported alongside absolute total pre-FSI cross sections~\cite{miniccqe}.

Recent MiniBooNE~\cite{miniccqe} and NOMAD~\cite{nomad} CCQE cross section measurements disagree under the relativistic Fermi gas model (Ref.~\cite{smithmoniz} is commonly used) by up to 30\% or more, despite using the same nuclear target ($^{12}$C). Surprisingly, the MiniBooNE CCQE cross section measurement (the larger of the two reported cross sections) surpasses the free neutron cross section. The discrepancy may be due to a CCQE multinucleon channel in which two correlated same-flavor nucleons are ejected~\cite{martini} and the experiments' varying levels of blindness to low energy protons coming out of the vertex or ``vertex activity". The channel may be the result of both short range correlations and 2 body currents between nucleons. Protons are not considered in the MiniBooNE CCQE measurement and protons with momentum less than $\sim$300 MeV/c are not reconstructed in the NOMAD analysis. MiniBooNE's complete insensitivity to the proton(s) is advantageous in the sense that the extracted cross section measurements do not rely as much on proton FSI modeling uncertainties~\cite{miniccqe}. However, the experiment is unable to measure a possible multinucleon excitation channel.

\begin{figure}[h!]
\centering
\begin{tabular}{c}
\hspace{-.3cm}
\epsfig{file=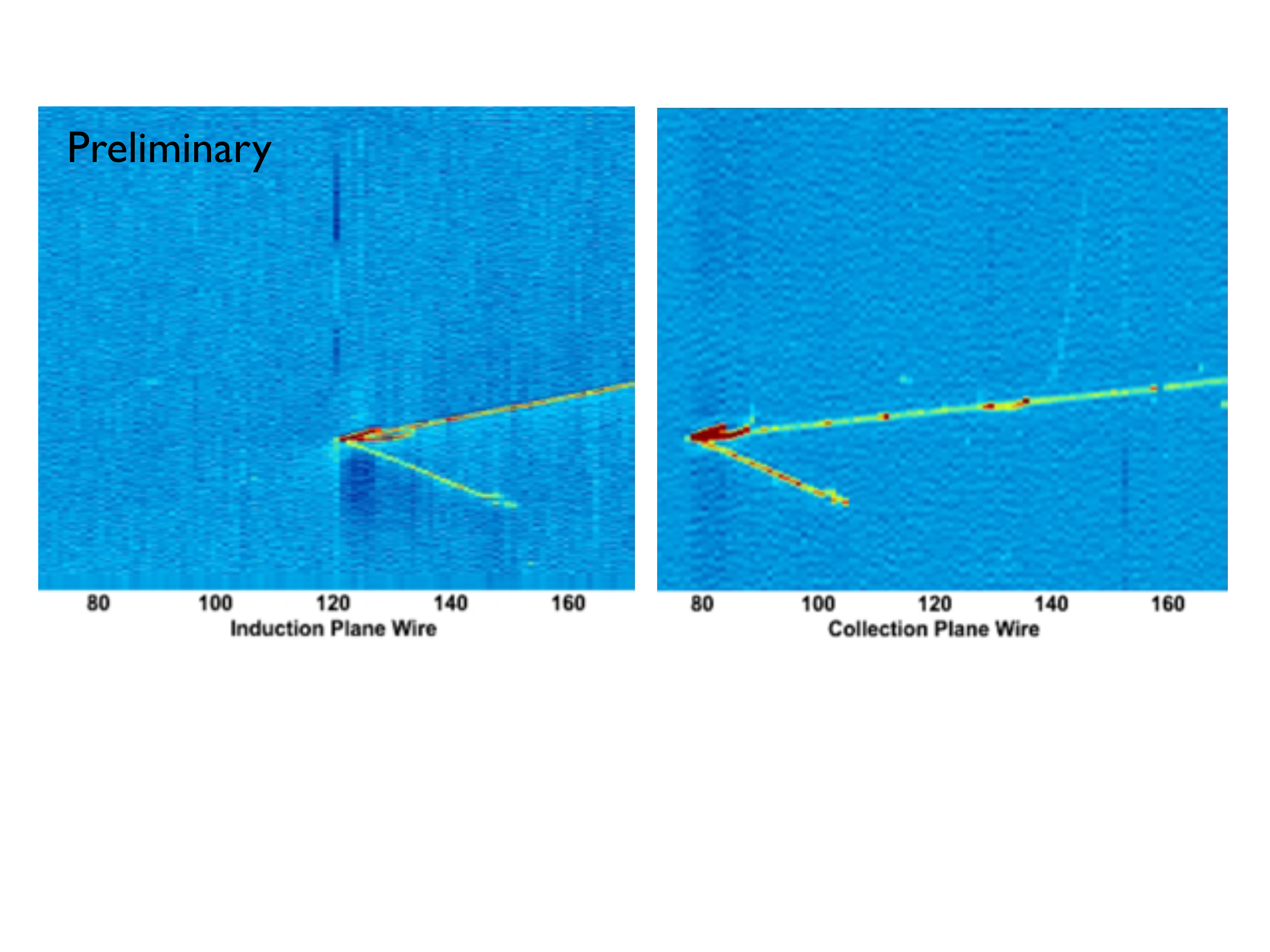,width=5.0in,angle=0}
\end{tabular}
\vspace{-2.5cm}
\caption{A zoomed-in neutrino CCQE candidate with clear evidence of multiple, highly ionizing protons and a muon (the topmost track) emanating from the vertex. The colors are indicative of the amount of energy deposited (blue$<$yellow$<$red). For a sense of scale, note that the wire spacing is 4~mm.}
\label{ccqe}
\end{figure}

Most neutrino detectors have a difficult time resolving vertex activity involving short tracks below Cherenkov threshold. ArgoNeuT, with its mm-scale resolution, three dimensional imaging, and few-10s of MeV energy threshold, will analyze the kinematics of vertex activity in CCQE interactions in an attempt to understand the multinucleon CCQE channel and final state interactions in general. Figure~\ref{ccqe} shows a CCQE interaction in ArgoNeuT with clear evidence of vertex activity through a set of $<$100~MeV protons. With the help of the downstream magnetized MINOS detector, ArgoNeuT will be able to compare neutrino and anti-neutrino CCQE events in terms of vertex activity. A multinucleon channel involving a 2 proton (neutron) final state before final-state-interactions in CCQE neutrino (anti-neutrino) events will be searched for. Comparing neutrino and anti-neutrino CCQE-like events may provide some sensitivity to this channel. Event characteristics such as energy deposited in a sphere of cm-scale radius around the vertex, number of protons, and proton kinematics in general and the differences between neutrino and anti-neutrino events in terms of these variables will be studied. Furthermore, the possibility of separating neutrino and anti-neutrino CCQE events with LArTPCs using vertex activity alone will be investigated. 

\section{Reconstructing the muon and finding the neutrino vertex}
The initial goals of ArgoNeuT's automated neutrino reconstruction software are to reconstruct the kinematics of the outgoing muon and find the neutrino interaction vertex. These steps are the key to beginning a CCQE analysis with ArgoNeuT. The muon reconstruction chain proceeds as follows:
\begin{itemize}
\item Event filter, to remove empty beam spills from the sample.
\item Deconvolution of the TPC and electronics response from the raw TPC-wire data, to produce standardized unipolar signal shapes.
\item Identification of hits above a threshold having a position, width, and projected ionization associated with each particle track.
\item DBSCAN (Density-Based Spatial Clustering of Applications with Noise)-based~\cite{dbscan} cluster finding, to group proximal hits together to form clusters and to identify noise hits.
\item Hough-transform-based~\cite{hough} cluster finding and track fitting, to find and fit line-like clusters/tracks. 
\item Three dimensional track matching with the wire signal and time information from both wire planes. 
\item Hit analysis and calorimetry, to characterize the hits that have been associated with long line-like tracks. 
\end{itemize}

\begin{figure}[h!]
\centering
\begin{tabular}{c}
\hspace{-.3cm}
\epsfig{file=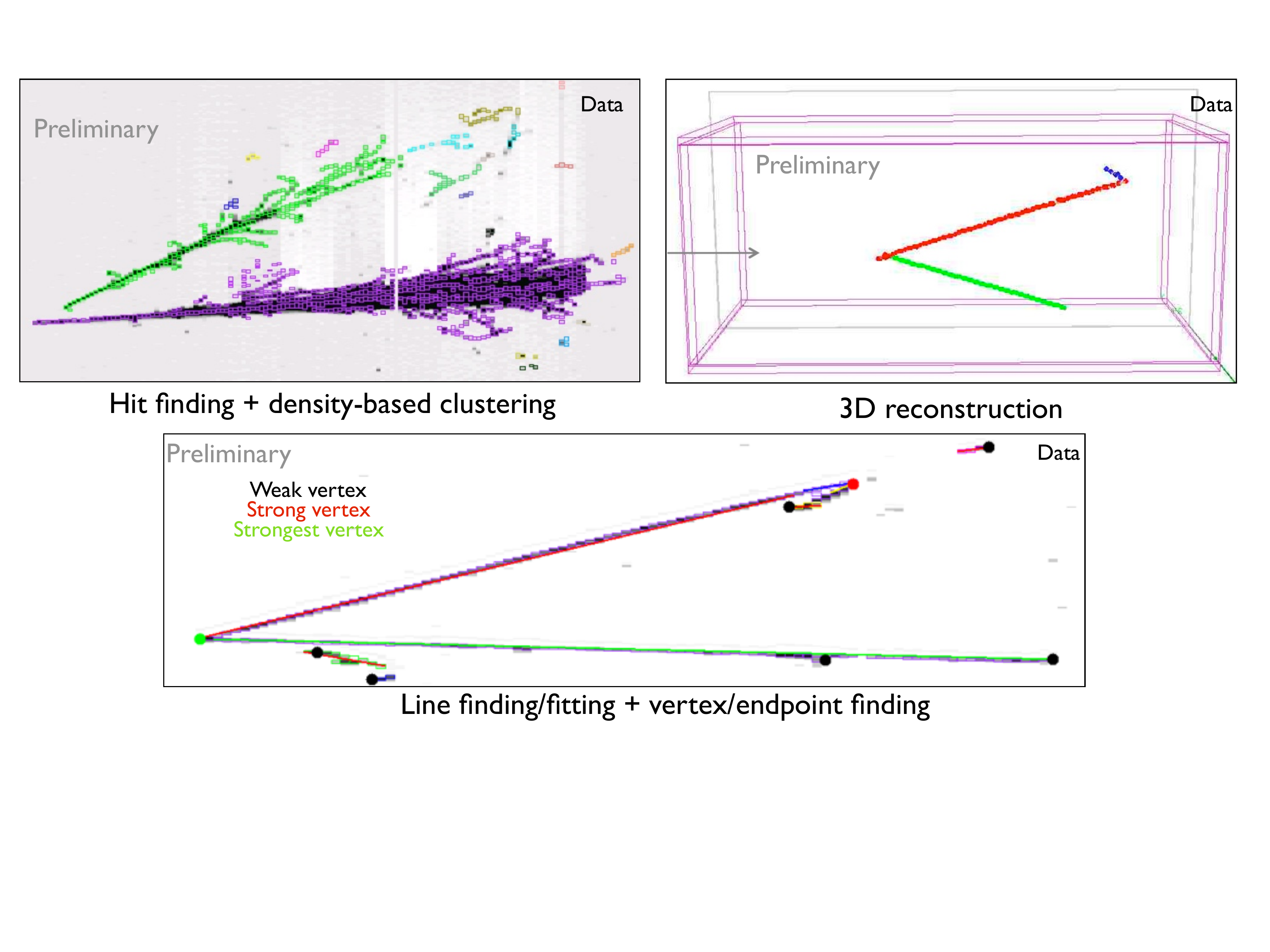,width=5.5in,angle=0}
\end{tabular}
\vspace{-2.5cm}
\caption{A few examples of the ArgoNeuT automated reconstruction software. (Top Left) A probable neutral current $\pi^{0}$ event ($\nu N\longrightarrow\nu N\pi^{0}$, $\pi^{0}\longrightarrow\gamma\gamma$) involving two energetic gamma-induced electromagnetic showers. The hit finding and density-based clustering algorithms show the hits and clusters associated with the event. (Top Right) A probable charged current resonant pion event ($\nu_{\mu}N\longrightarrow\mu^{-}N\pi^{+}$) reconstructed in three dimensions. (Bottom) The same event as seen on the induction plane with overlaid reconstructed line-like tracks and endpoints/vertices.}
\label{vertex}
\end{figure}

The vertex finding process takes hits that have been associated with DBSCAN clusters as an input and employs a Harris/Stephens-based~\cite{harris} algorithm to find the vertex and endpoints in the event. The algorithm considers the hits from a geometric point of view only with an implementation based on~Ref.~\cite{morgan}. These vertices/endpoints, after being given a ``strength", are matched with the endpoints of lines found with the Hough-transform-based line-finding/fitting algorithm to create ``strong vertices". An example of the reconstruction and endpoint/vertex finding algorithms at work can be seen in Figure~\ref{vertex}.

\section{Conclusion}
ArgoNeuT recently completed its $>$5 month physics run in the NuMI beamline at Fermilab. The first few thousand low energy neutrino and anti-neutrino events ever detected with a LArTPC are currently being analyzed with a fully automated software capable of reconstructing the muon and finding the neutrino interaction vertex. Among other channels, ArgoNeuT is investigating the CCQE cross section with a focus on the vertex activity associated with such events. With the unique ability to sign-select muons via the downstream MINOS near detector, ArgoNeuT will perform a direct neutrino/anti-neutrino comparison of low energy protons emanating from the interaction site. This measurement may be sensitive to a multinucleon excitation channel involving two same-flavor nucleons (protons, in the case of neutrino CCQE). This study will also seek to illuminate the influence of nuclear effects on the absolute and differential CCQE cross sections and on the experimental definition of CCQE itself.    

\section*{References}
\bibliographystyle{iopart-num}
\bibliography{./bibfile}

\end{document}